\documentclass[12pt,thmsa]{article}
 \usepackage{amsmath}

 \usepackage{amssymb}

\input{tcilatex}

\begin{document}

\baselineskip 17pt

\noindent \centerline {{\bf CHARGES AND FIELDS IN A CURRENT-CARRYING
WIRE}}

\vspace {13mm}

\centerline {\bf Dragan V Red\v zi\' c}

\noindent {\small Faculty of Physics, University of Belgrade, PO Box
368, 11001 Beograd, Serbia, Yugoslavia}\\ E-mail address:
redzic@ff.bg.ac.rs

\vspace {7mm}

\noindent {\bf Abstract}

\noindent Charges and fields in a straight, infinite, cylindrical
wire carrying a steady current are determined in the rest frames of
ions and electrons, starting from the standard assumption that the
net charge per unit length is zero in the lattice frame and taking
into account a self-induced pinch effect. The analysis presented
illustrates the mutual consistency of classical electromagnetism and
Special Relativity. Some consequences of the assumption that the net
charge per unit length is zero in the electrons frame are also
briefly discussed.

\bigskip

\noindent {\bf 1. Introduction}

As is well known, combining Coulomb's law, charge invariance and the
transformation law of a {\it pure} relativistic three-force [1,2],
one can derive the correct equation for the force with which a point
charge in uniform motion acts on any other point charge in {\it
arbitrary} motion, and thus recognize both the $\mbox {\pmb E}$ and
$\mbox {\pmb B}$ fields of a uniformly moving point charge and the
corresponding Lorentz force expression [3-5]. Thus one can prove
{\it indirectly}, without introducing general transformations for
$\mbox {\pmb E}$ and $\mbox {\pmb B}$, that the Lorentz force
expression, $\mbox {\pmb f}_L \equiv q\mbox {\pmb E} + q\mbox {\pmb
u}\times \mbox {\pmb B}$, transforms in the same way as the time
derivative of the relativistic momentum of a particle with {\it time
independent} mass, in the special case of $\mbox {\pmb E}$ and
$\mbox {\pmb B}$ due to a uniformly moving point charge. Following
the same line of reasoning, the so-called relativistic nature of the
magnetic field is often illustrated by discussing the force on a
charged particle outside a current-carrying wire, or the force
between two parallel current-carrying wires [5-8]. (Note that in the
latter case, contrary to the widespread opinion, both electrostatics
and magnetostatics need to be retained in the rest frame of
electrons, as van Kampen [9] pointed out, though not in an
irreproachable way [10,11].)

The above-mentioned arguments are intended to provide a short cut
towards classical electromagnetism via relativity. However, those
pedagogical vehicles, while ingenious, can be somewhat opaque to the
student. As can be seen, they are based on two mighty tacit
assumptions, namely, that Maxwell's equations are Lorentz covariant
and also that electromagnetic force on a point charge (which is the
Coulomb force in the right cases) transforms in the same way as
$d(m\mbox {\pmb u}/\sqrt {1 - u^2/c^2})/dt$ with $m =$ const;
unfortunately, to prove these assumptions without using four-vectors
is a real {\it tour de force} ([12-14], cf also [2]). Thus, the
familiar derivations that are aimed at reaching electromagnetism
from Coulomb's law (electrostatics) and relativity seem to be little
more than simple illustrations of the mutual consistency of
classical electromagnetism and Special Relativity (cf [15]).

An essential part in those pedagogical discussions, as well as in
related discussions of the transformation laws for $\mbox {\pmb E}$
and $\mbox {\pmb B}$ (cf, e.g., [6]), is played by a long conducting
wire carrying a steady current. The wire is usually modeled as
consisting of two superposed lines of charge, one moving and an
oppositely charged one at rest, extending along the $z$ axis. For
the convenience of the reader, we discuss briefly a typical example
that illustrates simply the consistency of electromagnetism and
relativity.

In the laboratory frame $S$, suppose there is a line of positive
charge at rest with constant linear charge density $\lambda_+$, and
a line of negative charge with density $\lambda_- = - \lambda_+$
moving to the right with a constant velocity $\mbox {\pmb v} =
v{\mbox {\pmb u}}_z$. Since the net charge density is always zero,
the electric field vanishes in the $S$ frame. However, due to a
current $I = |\lambda_-|v = \lambda_+v$ to the left, there is the
magnetic field whose direction is azimuthal around the line charges
with sense given by the right-hand rule; from Amp\` ere's law, $B =
\mu_0I/2\pi r$, where $r$ is the distance from the line charges. A
test charge $q$ placed at rest in $S$ remains at rest since there is
no electric force on it (the electric field is zero) and no magnetic
force (the test charge is at rest).

Consider now the same situation in the rest frame of the negative
charges, $S'$. In that frame, the positive charges move to the left
with velocity $-\mbox {\pmb v}$. The corresponding line charge
densities are $\lambda'_+ = \gamma \lambda_+$ and $\lambda'_- = -
\lambda_+/\gamma$, where $\gamma \equiv ({1 - v^2/c^2})^{-1/2}$, due
to the Lorentz contraction and charge invariance. Then $\lambda' =
\lambda'_+ + \lambda'_- = \lambda_+\gamma v^2/c^2$ and using Gauss'
law or otherwise ([5], [16,17], cf also [18]), we find that there is
a radial electric field in $S'$, pointing away from the line
charges, $E'_r = (1/2\pi \varepsilon_0r)\lambda_+ \gamma v^2/c^2$.
Also, due to the motion of the positive charges, there is an
azimuthal magnetic field in $S'$; using Amp\` ere's law we find that
$B' = \mu_0I'/2\pi r$, where $I' = \gamma \lambda_+ v = \gamma I$ is
the current in $S'$. The $\mbox {\pmb E}'$ and $\mbox {\pmb B}'$
fields can also be obtained directly, by Lorentz-transforming the
corresponding $\mbox {\pmb E}$ and $\mbox {\pmb B}$ fields in $S$.
The test charge $q$, at rest in $S$, moves uniformly with velocity
$-\mbox {\pmb v}$ relative to $S'$, consistent with the fact that
the corresponding Lorentz force, $q\mbox {\pmb E}' + q(-\mbox {\pmb
v})\times \mbox {\pmb B}'$, vanishes (and also consistent with the
transformation law of a pure relativistic three-force).

Some time ago, Peters [19] pointed out that the importance of
Special Relativity would be made more evident if in the above
example one used a more realistic model of a conducting wire. The
author considered an infinite cylindrical conductor of circular
cross section, whose material was idealized as consisting of the
lattice of immovable positive ions at rest in the lab frame $S$, and
an equal number of electrons which are free to move through the
lattice. The permittivity and permeability of such a material are
taken to be equal to those of free space, $\varepsilon_0$ and
$\mu_0$. When there is a steady current in the conductor, the
free-electrons have a nonzero average velocity $\mbox {\pmb v} =
v{\mbox {\pmb u}}_z$ in the rest frame of the lattice $S$ (``the
drift velocity"). In the $S'$ frame, the free-electrons are, on
average, at rest, and the lattice moves to the left with velocity
$-\mbox {\pmb v}$.

Now some simple questions arise. If one assumes that the infinite
wire carrying a steady current is electrically neutral in the
lattice frame $S$, is it possible to give an analysis of fields and
forces in the $S$ and $S'$ frames consistent with classical
electromagnetism and Special Relativity? Are the results obtained
for the more realistic model of the infinite conducting wire
analogous to those derived for the system consisting of superposed
lines of charge?

Peters came to a somewhat surprising conclusion that the bulk of the
conductor was neutral with respect to the rest frame of the
free-electrons $S'$ and {\it not} with respect to the lattice frame
$S$. Moreover, the author argued that the assumption of {\it
overall} neutrality of the wire in the $S$ frame led to a
contradiction. Namely, according to Peters, if the wire is neutral
in $S$ then there should exist a positive surface charge density on
the wire, generated by a self-induced pinch effect of the
free-electrons. A surface charge in $S$ implies the corresponding
surface charge in $S'$, due to charge invariance. However, Peters
claimed in [19] that no mechanism for generating a surface charge
existed in $S'$, and thus the assumption of overall neutrality in
$S$ was questionable. His argument was criticized by Hern\' andez
{\it et al} [20] who pointed out that there is a mechanism for
generating surface charge in $S'$ too. Gabuzda [21] proposed such a
mechanism and calculated the volume, surface and linear charge
densities in the $S$ and $S'$ frames.

It seems, however, that Peters' analysis of the problem contains a
difficulty with the concept of surface charge that went unnoticed by
the authors of References [20,21]. Also, Gabuzda's discussion [21]
seems to be based on a problematic starting assumption, which
appears occasionally in the literature. In this paper, we attempt to
give an analysis free from contradictions of charges and fields of
an infinite current-carrying wire modeled as in Reference [19], both
in the ions and electrons rest frames. The analysis leads to some
interesting insights and, hopefully, could be an intriguing reading
for the student of relativistic electrodynamics, together with our
recent contributions to the subject [2,22].

\medskip

\noindent {\bf 2. Charges and fields in the lattice frame}

\noindent Consider a straight, infinite, cylindrical conductor
having a circular cross section of radius $a$, the axis of the
conductor coinciding with the $z$ axis. The lattice consists of
uniformly distributed positive ions which are immovable; thus the
volume charge density of the lattice ions, $\rho_+$, is spatially
and temporally constant. When there is no current in the wire, the
drift velocity of the free-electrons, the electric field inside the
conductor, and the magnetic field everywhere are all zero. From
Gauss' law it follows that the volume charge density of the
free-electrons is $-\rho_+$. If the conductor is overall neutral,
then there is no surface charge over its surface, if it is
sufficiently far from other bodies.

Consider now the case when there is a steady current in the wire to
the left and assume, as is usually done, that all the free-electrons
have equal axial drift velocity $\mbox {\pmb v} = v{\mbox {\pmb
u}}_z$ to the right. The corresponding current density $\mbox {\pmb
J}$ is purely axial and, by symmetry, it can be expressed as a
function of a single variable $r$, denoting distance from the axis
of the conductor. Then, according to Ohm's law, there exists inside
the conductor an axial electrostatic field $\mbox {\pmb
E}_{\parallel} = E_{\parallel}(-{\mbox {\pmb u}}_z)$ which, by
symmetry, depends only on $r$ as well. The current produces a
magnetic field whose lines are circles around the conductor's axis,
with sense given by the right-hand rule. Consequently, there is a
magnetic force on the free-electrons directed inward. Thus, in a
steady configuration, there must exist also a transverse electric
field $\mbox {\pmb E}_{\perp}$ inside the wire, directed inward,
satisfying relation

\begin {equation}
\mbox {\pmb E}_{\perp} = -\mbox {\pmb v}\times \mbox {\pmb B}\, ,
\end {equation}

\noindent where $\mbox {\pmb B}$ is the magnetic flux density inside
the conductor due to the current in the conductor. Stress that
Equation (1) applies at the points inside the conductor where $\mbox
{\pmb J} \neq 0$, i. e. where the charge density of the
free-electrons is not zero. (Recall that an analogous situation
arises in the Hall effect, only in that case $\mbox {\pmb B}$ is not
due to the current itself but represents an externally applied
magnetic field.)

Now we shall determine the steady-state distribution of the
free-electrons inside the current-carrying wire. Using Gauss' law,
$\rho = \varepsilon_0 \nabla\cdot \mbox {\pmb E}$, and Equation (1),
we find that the net charge density inside the conductor, at the
points where Equation (1) applies, satisfies equation

\begin {equation}
\rho = \varepsilon_0 \nabla\cdot \mbox {\pmb E}_{\perp} =
\varepsilon_0 \mbox {\pmb v}\cdot (\nabla \times \mbox {\pmb B})\, ,
\end {equation}

\noindent since $\nabla\cdot \mbox {\pmb E}_{\parallel} = 0$ and
$\mbox {\pmb v}$ is constant. As $\rho$ and $\mbox {\pmb J}$ are
stationary, Amp\` ere's law applies

\begin {equation}
\nabla \times \mbox {\pmb B} = \mu_0\mbox {\pmb J} =
\mu_0\rho_-\mbox {\pmb v} \, ,
\end {equation}

\noindent where $\rho_-$ is the charge density of the free-electrons
in the steady state. Thus

\begin {equation}
\rho = \varepsilon_0 \mu_0\rho_-v^2 \equiv \rho_-v^2/c^2\, .
\end {equation}

\noindent Eventually, since $\rho = \rho_+ + \rho_-$, from Equation
(4) we get

\begin {equation}
\rho_- = -\rho_+\gamma^2 \, .
\end {equation}

\noindent Equation (5) implies that $\rho_-$ is spatially and
temporally constant, because $\rho_+$ is constant, and also that
$|\rho_-| > \rho_+$; thus, $\mbox {\pmb J}$ is constant too. A
simple calculus, starting from $\nabla \times \mbox {\pmb E} = 0$
and Equation (1), reveals that $\mbox {\pmb E}_{\parallel}$ is
constant as well, which, in combination with previous results and
Ohm's law, implies that all the free-electrons have equal
mobilities. (Note that in the problem considered, the usual
statement of Ohm's law, $\mbox {\pmb J} = \sigma \mbox {\pmb E}$,
where $\sigma$ is the conductivity of the material, need to be
replaced by a more general form, $\mbox {\pmb J} = \sigma (\mbox
{\pmb E} + \mbox {\pmb v}\times \mbox {\pmb B})$, which reduces to
$\mbox {\pmb J} = \sigma \mbox {\pmb E}_{\parallel}$.) Equations (4)
and (5) were obtained using Gauss' law and Amp\` ere's law in their
integral forms by Matzek and Russell [23]; the differential approach
used above is adapted from Rosser [24].

Greater density of the free-electrons when they are in motion was
explained by previous authors [23,19,21] by a self-induced pinch
effect of the electrons. Namely, during the transient build-up of
the current in the wire, a magnetic pinch will develop as the
electrons in the wire are macroscopically accelerated (i. e. as
their mean velocity varies from zero up to the steady drift velocity
$\mbox {\pmb v}$). Consequently, when equilibrium is reached, the
free-electrons are concentrated towards the axis of the wire,
leaving a region at the conductor's surface which is swept clear of
the electrons by this pinch effect, as was pointed out in [23].
Thus, there remains a layer of positive ions at the surface, whose
thickness $\delta$ can be determined from the assumption that the
wire is neutral also when it carries a steady current. Is that
assumption plausible?

We are faced here with a tricky problem of how is a steady current
established in the conducting wire. It seems that there are two
solutions to the problem, depending on initial conditions; this will
be discussed in some detail in the last section. Now, as is usually
done for teaching purposes, we shall find the thickness of the
``naked" layer of positive ions starting from the standard
assumption that the current in the wire is established in such a way
that the net charge per unit length of the wire is zero in the
lattice frame. (Recall that the standard argument in ``current
without pinch" discussions claims that ``the only difference between
a wire carrying a current and a wire not carrying a current is the
existence of a drift velocity for the electrons. The mean distance
between the electrons remains unaffected as measured in the
laboratory frame", [5, p 259], cf also [18], and the last section.)

That assumption obviously implies

\begin {equation}
\rho_+ a^2\pi l + \rho_-(a - \delta)^2 \pi l = 0\, ,
\end {equation}

\noindent where $l$ is the length of a segment of the conductor
between two parallel planes which are perpendicular to its axis.
From Equations (5) and (6) one has

\begin {equation}
a - \delta = a\sqrt {1 - v^2/c^2} \equiv a/\gamma\, .
\end {equation}

The electron drift velocities are typically fractions of millimeter
per second so that the ratio $v/c$ is about $10^{-12}$, and the
thickness of the naked layer $\delta$ is ``almost inconceivably
small". It is therefore tempting to assume that the thickness
$\delta$ may be neglected in calculations for all reasonable values
of $a$ and introduce the corresponding surface charge density. So
some authors [19,21] take that, in a steady configuration, Equations
(4) and (5) are valid for all $r < a$ and calculate the
corresponding surface charge density $\varsigma$ from the condition

$$
(\rho_+ + \rho_-)a^2\pi l + \varsigma 2\pi al = 0\, ,
$$

\noindent which gives $\varsigma = |\rho|a/2$. However, in this way
extra free-electrons of charge $\rho_- [a^2 - (a - \delta)^2] \pi l$
and an opposite extra positive charge $\varsigma 2\pi al$, have been
created {\it ex nihilo} and added to the existing charge
distribution in the wire segment, described by Equations (4)-(7). As
can be seen, the amounts of these fictitious extra charges are
comparable to those of the existing charges in the wire segment.
Consequently, it appears that the thickness of the naked layer
$\delta$ must not be neglected and thus the concept of surface
charge should be discarded in this case. It can be methodologically
dangerous to neglect $v^2/c^2$ at one point of analysis, and to
retain the same quantity elsewhere. (True, there is a way to save
the surface charge concept, altering {\it distances} but not {\it
charges}. Thus, one could argue that the correct surface charge
density, $\varsigma^*$, is found from the condition $\rho_+ [a^2 -
(a - \delta)^2] \pi l = \varsigma^* 2\pi al$, and the volume charge
densities $\rho_+$ and $\rho_-$ should be modified accordingly. For
example, the corresponding positive volume charge density would be
given by $\rho_+^* \pi a^2l = \rho_+ (a - \delta)^2\pi l$. However,
the above procedure gives an approximate solution to the problem
which is, to say the least, more complicated than the correct one.)

From the preceding considerations it follows that in the present
analysis we should retain $\delta$ everywhere and thus, somewhat
surprisingly, only volume and linear densities of charge should be
used.

It is now simple to determine the electric and magnetic fields in
the $S$ frame using $\nabla \times \mbox {\pmb E} = 0$, Gauss' law,
Amp\` ere's law, and Ohm's law, and taking into account that the
current $I$ in the conductor is uniformly distributed inside the
circle of radius $r = a - \delta$. For the sake of completeness, we
present the final results for $\mbox {\pmb E}$ and $\mbox {\pmb B}$.

The magnitude of the magnetic flux density is given by

\begin {equation}
B = \left \{  \begin {array}{ll} \mu_0Ir/2\pi (a - \delta)^2\,, & r \leq a - \delta\, ,\\
    \mu_0I/2\pi r\,, & r > a - \delta\, .
    \end {array}\right.
\end {equation}

The axial electric field is constant {\it everywhere}

\begin {equation}
\mbox {\pmb E}_{\parallel} = - I\mbox {\pmb u}_z/\pi (a -
\delta)^2\sigma \, ,
\end {equation}

\noindent and the transverse (radial) electric field can be
expressed as

\begin {equation}
\mbox {\pmb E}_{\perp} = \left \{  \begin {array}{cl} -\mu_0Ivr\mbox
{\pmb u}_r/2\pi (a - \delta)^2\,,
& r \leq a - \delta\, ,\\
    -\mu_0Iv(a^2 - r^2)\mbox {\pmb u}_r/2\pi ra^2(v^2/c^2)\,, & a - \delta \leq r \leq a\, ,\\
    0\,, & r > a\, ,
    \end {array}\right.
\end {equation}

\noindent using relation

\begin {equation}
I = \rho_+\pi a^2v\, .
\end {equation}

If there is a test charge $q$ near the wire which is instantaneously
at rest relative to the $S$ frame, then the force $\mbox {\pmb F} =
q\mbox {\pmb E}_{\parallel}$ acts on the charge at that moment.
Next, if $q > 0$, the charge would deflect towards the wire due to
the combined action of $\mbox {\pmb E}_{\parallel}$ and $\mbox {\pmb
B}$.

\medskip

\noindent {\bf 3. Charges and fields in the free-electrons frame}

\noindent Consider now the same current-carrying wire in the
steady-state configuration as observed in the $S'$ frame, which
moves with velocity $\mbox {\pmb v} = v\mbox {\pmb u}_z$ relative to
the $S$ frame. In $S'$, the free-electrons are, on average, at rest,
and the positive lattice ions move to the left with velocity $-\mbox
{\pmb v}$. One easily finds that $\rho_+' = \gamma \rho_+$ and
$\rho_-' = \rho_-/\gamma = -\gamma \rho_+$, due to the Lorentz
contraction, charge invariance and Equation (5). Since transverse
lengths are Lorentz invariant, the radius of the wire is still $a$,
and the free-electrons are still inside the cylinder of radius $a -
\delta$. Thus, the bulk of the conductor ($r < a - \delta$) is
neutral in the $S'$ frame! However, the net charge per unit length
of the conductor, $\lambda'$, is not zero, due to the naked layer of
moving positive ions ($a - \delta < r < a)$. Obviously,

\begin {equation}
\lambda' = \rho'_+\pi[a^2 - (a - \delta)^2] = \gamma \rho_+ \pi
a^2v^2/c^2\,,
\end {equation}

\noindent using relation $\rho_+' = \gamma \rho_+$, Equation (7) and
identity $\gamma^2 - 1 \equiv \gamma^2 v^2/c^2$. Equation (12) is
consistent with the corresponding equation for the simple system of
two superposed lines of charge, discussed in the Introduction, since
$\rho_+ \pi a^2 \equiv \lambda_+$. (Note that the result for
$\lambda'$ derived in Reference [21], Equation (5) in [21], which in
our notation reads $\lambda' = \gamma^3 \rho_+ \pi a^2v^2/c^2$, is
different from our result (12) due to the contribution of the
fictitious surface charges introduced in [21].)

The thickness $\delta$ of the naked positive layer in $S'$ is
deduced above from Lorentz invariance of transverse lengths. The
same  result is obtained by applying charge conservation in $S'$.
Namely, in the ``initial", ``current without pinch", situation in
$S'$, the lattice shrinks along the direction of motion whereas the
axial expansion of the free-electrons occurs, due to the Lorentz
contraction, as compared with the corresponding ``current without
pinch" situation as observed in the $S$ frame. Since the ions and
electrons charge densities in $S$ are then $\rho_+$ and $-\rho_+$
{\it everywhere}, the corresponding $S'$ densities are
$\rho_+\gamma$ and $-\rho_+/\gamma$ throughout the wire.
Consequently, there is a radial electric field directed outward in
the wire in $S'$ due to an excess of positive charge, and the
free-electrons move inward until the net charge density is zero in
the coaxial cylindrical region enriched by the electrons and the
radial electric field vanishes, as was suggested by Matzek and
Russell [23]. Thus, charge conservation applied to a wire segment of
length $l$ in $S'$ gives

\begin {equation}
[\rho_+\gamma + (-\rho_+/\gamma)] a^2\pi l = \rho_+\gamma a^2\pi l +
(-\rho_+\gamma)(a - \delta)^2 \pi l\, ,
\end {equation}

\noindent wherefrom we find that $\delta$ satisfies Equation (7), as
it should.

The fact that the net linear charge density is zero in $S$ and
nonzero in $S'$ might appear at first sight contradictory to charge
invariance; but the paradox is only apparent. Namely, as can be
seen, for a well-defined wire segment in the lattice frame $S$ (the
positive ions at rest and an identical number of moving
free-electrons taking up the same length $d$ in $S$ at one instant
of the $S$ time), there is no corresponding {\it wire segment} in
$S'$ because these ions and electrons take up different lengths in
$S'$ at a certain instant of the $S'$ time, $d/\gamma$ and
$d\gamma$, respectively. In other words, wire segments cannot be
reified, the ions and the corresponding electrons must be treated
separately when they are in motion with respect to each other, as
Webster [25] pointed out (cf also [10,11]). The above resolution to
the paradox seems to be more satisfactory than that proposed in
Reference [18], involving another infinite wire.

The above argument implies the following $S$- and $S'$-scenarios. In
$S$, in the original ``no current" state, assuming overall
neutrality, the ions and electrons volume charge densities are
$\rho_+$ and $-\rho_+$. Next, under the action of the axial electric
field $\mbox {\pmb E}_{\parallel} = E_{\parallel}(-{\mbox {\pmb
u}}_z)$, the free-electrons acquire the drift velocity $\mbox {\pmb
v} = v{\mbox {\pmb u}}_z$ while their charge density is still
$-\rho_+$ everywhere in that ``current without pinch" state [5,18].
Eventually, the pinch effect develops under the action of the
magnetic field of the current itself, charge separation occurs
producing a transverse electric field, naked positive layer and
final electrons charge density $-\rho_+\gamma^2$. On the other hand,
in $S'$, in ``no current" state, the corresponding charge densities
are $\rho_+\gamma$ and $-\rho_+\gamma$ everywhere, both ions and
electrons (on average) travel to the left with velocity $-\mbox
{\pmb v}$, producing two {\it convection} currents equal in
magnitude but of opposite directions. Next, under the action of an
axial electric field $\mbox {\pmb E}'_{\parallel}$ to the left, the
electrons, on average, stop and remain at rest, now with charge
density $-\rho_+/\gamma$ throughout the wire, and thus only the ions
current to the left survives,

\begin {equation}
I' = \rho'_+\pi a^2v = \gamma I\, ,
\end {equation}

\noindent using Equation (11). (The analogous asymmetry between the
$S$- and $S'$-descriptions is found in the well-known
``thread-between-spaceships" relativistic problem [26], cf also
[27-30].) Eventually, the pinch effect develops under the action of
the radial electric field directed outward, leading to final
electrons charge density $-\rho_+\gamma$ in the cylindrical region
$(r < a - \delta)$ and thus to vanishing of the radial electric
field inside the cylinder.

It should be stressed that the above distinction between ``current
without pinch" and ``current with pinch" stages is somewhat
simplistic; a more realistic scenario should take into account that
the pinch effect develops at the same time as the current
establishes in the wire. However, conceptual traps lurk even in the
simplistic scenario.

Now we shall briefly discuss the electric and magnetic fields in
$S'$.

In $S$, the constant axial electric field $\mbox {\pmb
E}_{\parallel} = E_{\parallel}(-{\mbox {\pmb u}}_z)$ given by
Equation (9) is needed to produce and keep the steady (conduction)
current $I$ to the left due to the free-electrons with drift
velocity $\mbox {\pmb v} = v{\mbox {\pmb u}}_z$ to the right. On the
other hand, in $S'$, an axial electric field $\mbox {\pmb
E}'_{\parallel}$ is needed to annihilate the initial convection
current of the free-electrons (due to their motion with velocity $-
\mbox {\pmb v}$ to the left), and keep them, on average, at rest,
despite their ``home-lattice" remains in uniform motion with
velocity $- \mbox {\pmb v}$. Symmetry suggests that {\it everywhere}

\begin {equation}
\mbox {\pmb E}'_{\parallel} = \mbox {\pmb E}_{\parallel}\, .
\end {equation}

\noindent (Recall, however, that one should be careful with
symmetries in Special Relativity [31,32], [27,28].) Equation (15) is
consistent with the transformation law for $\mbox {\pmb E}$ and
$\mbox {\pmb B}$

\begin {equation}
\begin {array}{ll}\mbox {\pmb E}_\parallel = \mbox {\pmb E}'_\parallel\, ,
\qquad & \mbox {\pmb E}_\perp = \gamma [\mbox {\pmb E}'_\perp -
\mbox {\pmb
v} \times \mbox {\pmb B}']\, , \\
 & \\
\mbox {\pmb B}_\parallel = \mbox {\pmb B}'_\parallel\, , \qquad &
\mbox {\pmb B}_\perp = \gamma [\mbox {\pmb B}'_\perp + (1/c^2) \mbox
{\pmb v} \times \mbox {\pmb E}']\, ,
\end {array}
\end {equation}

\noindent where $\parallel$ and $\perp$ denote the field components
parallel and normal to $\mbox {\pmb v}$, respectively [33,12]. Also,
as can be seen, Equation (15) is consistent with contraption for
producing $\mbox {\pmb E}_\parallel$ in $S$, consisting of two
parallel charged conducting planes at rest in $S$ which are
perpendicular to the wire [34,35].

\noindent [Note that a related problem of an infinitely long
cylindrical wire carrying a steady current with cylindrically
symmetric return path is usually considered without taking into
account the self-induced pinch-effect. In that approximation, a
correct analytic solution implies that there is a distribution of
charge over the surface of the wire whose density is a {\it linear}
function of the axial coordinate $z$, cf [36] and also [37].
However, as Sommerfeld pointed out, ``the zero point of the charge
remains undetermined since the point $z = 0$ can be fixed
arbitrarily. We may eventually identify it with the ``center" of the
wire, which, for infinite length, also remains indefinite" [36]. In
our opinion, Sommerfeld's statement implies that the linear function
solution for the surface charge density is {\it physically}
meaningless. As can be seen, the problem persists even if the pinch
effect is taken into account.]

Eventually, using Equations (9) and (15), Amp\` ere's law and Gauss'
law we find $\mbox {\pmb E}'_\perp$ and $\mbox {\pmb B}'$.

The magnitude of $\mbox {\pmb B}'$ is given by

\begin {equation}
B' = \left \{  \begin {array}{ll} \mu_0\gamma Ir/2\pi a^2\,, & r \leq a, ,\\
    \mu_0\gamma I/2\pi r\,, & r \geq a \, ,
    \end {array}\right.
\end {equation}

\noindent and the transverse (radial) electric field can be
expressed as

\begin {equation}
\mbox {\pmb E}'_{\perp} = \left \{  \begin {array}{cl} 0\,,
& r \leq a - \delta\, ,\\
   \gamma I[r^2 - (a - \delta)^2]\mbox {\pmb u}_r/2\pi \varepsilon_0 ra^2v\,, & a - \delta \leq r \leq a\, ,\\
    \mu_0 \gamma Iv\mbox {\pmb u}_r/2\pi r\,, & r \geq a\, ,
    \end {array}\right.
\end {equation}

\noindent using relation $\rho'_+ = \gamma \rho_+$ and Equations (7)
and (11).

It can be easily verified that expressions (8)-(10) and (15), (17)
and (18) for the electric and magnetic fields in the $S$ and $S'$
frames are consistent with the transformation law (16), as they
should be. Also, the test charge $q > 0$ which is momentarily at
rest in $S$, and has the instantaneous velocity $-\mbox {\pmb v}$ at
the same space-time point in $S'$,  would subsequently deflect
towards the wire, as observed in $S'$ too.

\medskip

\noindent {\bf 2. Concluding comments}

\noindent The above analysis of charges and fields of an infinitely
long conducting wire carrying a steady current in the lattice frame
and the free-electrons frame provides another illustration of the
mutual consistency of classical electromagnetism and Special
Relativity. It should be stressed that our conclusions are
essentially based on a purely mathematical fact that Maxwell's
equations and equation of motion of a charged particle in the
electromagnetic field

$$
\frac {d}{dt}\left (\frac {m\mbox {\pmb u}}{\sqrt {1 -
u^2/c^2}}\right ) = q(\mbox {\pmb E} + \mbox {\pmb u}\times \mbox
{\pmb B})\, ,
$$

\noindent where the mass $m$ is a time independent Lorentz scalar,
are covariant with respect to the Lorentz transformation, regardless
of the value of the ratio $v/c$ appearing in it. Thus French's
exclamation near the end of his excellent book: ``Who says
relativity is important only for velocities comparable to that of
light?" [5], in the context of an analysis similar to ours, seems to
be somewhat exaggerated. However, this does not mean that we in any
way doubt Special Relativity or its importance even at ``creeping"
velocities (cf, e.g., [31], [38-40]).

Also, the above analysis leads to two somewhat surprising insights:
first, the concept of surface charge must not be introduced for
every extremely thin layer and, second, wire segments in our model
cannot be reified. While the first insight is not related directly
to Special Relativity, the second one is in opposition to our
Galilean instincts, showing clearly why it is hard to acquire a
relativistic mentality. In addition to these conceptual traps, it
should be pointed out that some of our starting assumptions appear
to be questionable.

First point, in a copper wire there is about $10^{23}$ Cu$^+$
ions/cm$^3$, and for $v = 1$mm/s using Equation (5) we obtain that
roughly 1 more electrons than protons is found in each cubic
centimeter of the copper wire. For a wire with a cross-sectional
area of about 1mm$^2$, this amounts to one electron per meter of the
wire, which presumably suggests breakdown of the continuum model in
our case. (Note that the same trouble with second-order charges
appears in a recent analysis of a rotating conducting sphere [41].)

Another difficult point is the standard assumption (usually
introduced by {\it fiat}, cf, e.g. [7]) that the net charge per unit
length of an infinite cylindrical wire carrying a steady current is
zero in the lattice frame. (The assumption is closely related to the
Clausius postulate, ``a closed constant current in a stationary
conductor exerts no force on stationary electricity" [42,43], [31],
[38].) Zapolsky [18] gave a relativistic justification of the
standard assumption in the framework of an elementary but nontrivial
``procession" model of conductor carrying a steady current for both
an infinite straight conducting wire and a thin circular wire. While
his argument appears to be correct in the latter case, it seems to
be problematic in the case of the infinite wire since the author
starts from the assumption that a constant axial electric field is
``turned on" along the infinite wire at the moment $t = 0$. While
that assumption is often found in textbooks (cf, e. g., [44]), it is
clearly incompatible with Special Relativity. (The ``turning-on" of
the external electric field is a rather intricate process. For some
insights into the analogous ``turning-off" physics cf, e.g., [45].)
Stress, however, that Zapolsky is right that spacing between two
electrons which are accelerated, starting simultaneously from rest,
in a constant electric field always remains equal to their initial
separation, all with respect to the $S$ frame [18].

On the other hand, some authors [46] seem to imply that the spacing
between electrons contracts as they are accelerated and thus deduce
that the electronic charge distribution contracts as it is
accelerated up to the drift velocity $v$ during the transient
buildup of the current, all with respect to $S$, invoking the
current sinks and sources (i. e. batteries) in order to bring the
wire back into neutrality. However, since the starting assumption is
wrong (as Zapolsky [18] pointed out), the argument is inconclusive.
Note that the same problematic assumption - that of the Lorentz
contraction of the electronic charge distribution as compared to its
initial (zero drift velocity) distribution - is implicit in a recent
paper by Brill {\it et al} [47]. Namely, the authors claim that a
straight, infinite, current-carrying wire, modeled as consisting of
two infinite lines of charge (the ions at rest in $S$ and the
electrons at rest in $S'$), is electrically neutral neither in $S$
nor in $S'$ but in the ``middle frame" for the $S$ and $S'$ frames
(whose speed is $(1 - \gamma^{-1})c^2/v$ relative to both frames, as
a simple calculation shows). As can be seen, their claim would be
correct only if, in $S$, distances between adjacent positive ions at
rest and adjacent electrons moving at the drift speed $v$ were, say,
$d$ and $d\sqrt {1 - v^2/c^2}$, respectively (the respective
distances as measured in $S'$ would be $d\sqrt {1 - v^2/c^2}$ and
$d$). Recall that the same assumption of electrical neutrality in
the {\it middle frame} was essentially tacitly used in the {\it
first} edition of Purcell's book [7], and also in [8]. The same
assumption was recently revived in the case of a cylindrical wire by
Folman [48].\footnote [1] {It seems that the possibility that the
wire is neutral in the middle frame for $S$ and $S'$, involving the
Lorentz contraction of the electronic charge distribution relative
to the lattice frame, must be ruled out on the basis that the free
electrons are not connected to rigidly moving rods.}

One last point. As the above argument shows, the standard assumption
that the net charge per unit length is zero in the lattice frame $S$
is essentially based on Zapolsky's insight that spacing between two
electrons does not change with time if the electrons are accelerated
starting simultaneously from rest in a constant electric field, all
with respect to $S$. (As observed in $S'$, {\it decelerations} of
the two electrons neither start nor stop simultaneously, which
accounts for increase in their separation with respect to $S'$.)
What if, instead of the standard assumption, we assume that the net
charge per unit length is zero in the free-electrons frame $S'$ not
only when there is no current in the wire but also when it carries a
steady current? Obviously, that would require that we somehow manage
to decelerate simultaneously until they stop the two electrons with
respect to $S'$, and thus their separation would not change with
time in $S'$. Then the ions and electrons volume charge densities
would be $\rho_+\gamma$ and $-\rho_+\gamma$ throughout the wire also
when there is current in it (i. e. when the free-electrons stop and
remain at rest). No pinch develops because there is no transverse
electric force (no transverse electric field) and no magnetic force
(the electrons are at rest). The corresponding charge densities in
$S$ would be $\rho_+$ and $-\rho_+\gamma^2$ throughout the wire; as
can be seen, no pinch develops in $S$ either, because the transverse
electric force and the magnetic force on the moving electrons cancel
each other.

To summarize, it appears that the problem of in what frame is an
infinite current-carrying wire neutral does not have a unique
solution. It can be neutral either in the lattice frame $S$ or in
the electrons frame $S'$, depending on the way the current is
established in the wire (by a simultaneous acceleration of the
electrons relative to $S$, or by a simultaneous deceleration of the
electrons relative to $S'$, respectively). Both cases are consistent
with Special Relativity, since the corresponding equilibrium
situations in the $S$ and $S'$ frames are described on the basis of
classical electromagnetism, which is Lorentz covariant. Our simple
results for equilibrium charge and current distributions satisfy
Maxwell's equations. Thus, perhaps, the above analysis is not devoid
of physical sense.

\medskip

\noindent {\bf Acknowledgment}

\noindent The author would like to acknowledge support of the
Ministry of Science and Education of the Republic of Serbia (project
No. 171029), for this work.

\newpage

\noindent {\bf References}

\medskip

\noindent [1] Rindler, W.,  {\it Introduction to Special
Relativity}, 2nd ed., 90-93, Clarendon, Oxford, 1991.

\noindent [2] Red\v zi\' c, D. V.,  D. M. Davidovi\' c and M. D.
Red\v zi\' c, ``Derivations of relativistic force transformation
equations," {\it J. Electro. Waves Appl.}, Vol. 25, 1146-1155, 2011.

\noindent [3] Tolman, R. C., ``Non-Newtonian mechanics.-Some
transformation equations," {\it Phil. Mag. S. 6}, Vol. 25, 150-157,
1913.

\noindent [4] Rosser, W. G. V., ``The electric and magnetic fields
of a charge moving with uniform velocity" {\it Contemp. Physics},
Vol. 1, 453-466, 1960.

\noindent [5] French, A. P., {\it Special Relativity}, Nelson,
London, 1968.

\noindent [6] Resnick, R., {\it Introduction to Special Relativity},
Wiley, New York, 1968.

\noindent [7] Purcell, E. M., {\it Electricity and Magnetism} 2nd
ed, McGraw-Hill, New York, 1985.

\noindent [8] Griffiths, D. J., {\it Introduction to
Electrodynamics}, 3rd ed., Prentice-Hall, Upper Saddle River, NJ,
1999.

\noindent [9] van Kampen, P.,  ``Lorentz contraction and
current-carrying wires," {\it Eur. J. Phys.}, Vol. 29, 879-883,
2008.

\noindent [10] Red\v zi\' c, D. V.,  ``Comment on `Lorentz
contraction and current-carrying wires'," {\it Eur. J. Phys.}, Vol.
31, L25-L27, 2010.

\noindent [11] van Kampen, P.,  ``Reply to `Comment on ``Lorentz
contraction and current-carrying wires"'," {\it Eur. J. Phys.}, Vol.
31, L29-L30, 2010

\noindent [12] Rosser, W. G. V.,  {\it An Introduction to the Theory
of Relativity}, 303-310, Butterworth, London, 1964.

\noindent [13] Schwartz, H. M., ``Einstein's comprehensive 1907
essay on relativity, part II," {\it Am. J. Phys.}, Vol. 45, 811-817,
1977.

\noindent [14] Jefimenko, O. D., ``Derivation of relativistic force
transformation equations from Lorentz force law," {\it Am. J.
Phys.}, Vol. 64, 618-620, 1996.

\noindent [15] Jefimenko, O. D., ``Is magnetic field due to an
electric current a relativistic effect? ," {\it Eur. J. Phys.}, Vol.
17, 180-182, 1996.

\noindent [16] Jefimenko, O. D., ``Retardation and relativity: The
case of a moving line charge," {\it Am. J. Phys.}, Vol. 63, 454-459,
1995.

\noindent [17] Rosser, W. G. V., ``Comment on `Retardation and
relativity: The case of a moving line of charge,' by Oleg D.
Jefimenko [Am. J. Phys. 63(5), 454-459 (1995)]," {\it Am. J. Phys.},
Vol. 64, 1202-1203, 1996.

\noindent [18]  Zapolsky, H. S., ``On electric fields produced by
steady currents," {\it Am. J. Phys.}, Vol. 56, 1137-1141, 1988.

\noindent [19]  Peters, P. C., ``In what frame is a current-carrying
conductor neutral," {\it Am. J. Phys.}, Vol. 53, 1165-1169, 1985.

\noindent [20] Hern\' andez, A., M. A. Valle and J. M.
Aguirregabiria, ``Comment on `In what frame is a current-carrying
conductor neutral'," {\it Am. J. Phys.}, Vol. 56, 91, 1988.

\noindent [21] Gabuzda, D. C., ``The charge densities in a
current-carrying wire," {\it Am. J. Phys.}, Vol. 61, 360-362, 1993.

\noindent [22] Red\v zi\' c, D. V.,  M. S. A. Eldakli and M. D.
Red\v zi\' c, ``An extension of the Kelvin image theory to the
conducting Heaviside ellipsoid," {\it Progress In Electromagnetics
Research M}, Vol. 18, 233-246, 2011.

\noindent [23] Matzek, M. A. and B. R. Russell, ``On the transverse
electric field within a conductor carrying a steady current," {\it
Am. J. Phys.}, Vol. 36, 905-907, 1968.

\noindent [24] Rosser, W. G. V., ``Magnitudes of surface charge
distributions associated with electric current flow," {\it Am. J.
Phys.}, Vol. 38, 265-266, 1970.

\noindent [25] Webster,D. L., ``Relativity and parallel wires," {\it
Am. J. Phys.}, Vol. 29, 841-844, 1961.

\noindent [26] Red\v zi\' c, D. V.,  ``Note on Dewan-Beran-Bell's
spaceship problem," {\it Eur. J. Phys.}, Vol. 29, N11-N19, 2008.

\noindent [27] Peregoudov, D. V.,  ``Comment on `Note on
Dewan-Beran-Bell's spaceship problem'," {\it Eur. J. Phys.}, Vol.
30, L3-L5, 2009.

\noindent [28] Red\v zi\' c, D. V.,  ``Reply to `Comment on ``Note
on Dewan-Beran-Bell's spaceship problem"'," {\it Eur. J. Phys.},
Vol. 30, L7-L9, 2009.

\noindent [29] Red\v zi\' c, D. V., ``Relativistic length agony
continued" arXiv: 1005.4623, 2010.

\noindent [30] Podosenov, S. A.,  J. Foukzon and A. A. Potapov, ``A
study of the motion of a relativistic continuous medium," {\it
Gravitation and Cosmology}, Vol. 16, 307-312, 2010.

\noindent [31] Bartocci, U.,  and M. M. Capria, `Symmetries and
asymmetries in classical and relativistic electrodynamics," {\it
Found. Phys.}, Vol. 21, 787-801, 1991.

\noindent [32] Red\v zi\' c, D. V., ``Comment on `Some remarks on
classical electromagnetism and the principle of relativity,' by U
Bartocci and M Mamone Capria," {\it Am. J. Phys.}, Vol. 61, 1149,
1993.

\noindent [33] Einstein, A., ``Zur Elektrodynamik bewegter
K\"{o}rper," {\it Ann. Phys., Lpz.}, Vol. 17, 891-921, 1905.

\noindent [34] Marcuse, D., ``Comment on `The electric field outside
a long straight wire carrying a steady current'," {\it Am. J.
Phys.}, Vol. 38, 935-936, 1970.

\noindent [35] Pugh, E. M., ``Poynting vectors with steady
currents," {\it Am. J. Phys.}, Vol. 39, 837-838, 1971.

\noindent [36] Sommerfeld, A.,  {\it Electrodynamics}, translated by
E G Ramberg, Academic Press, New York, NY, 1952.

\noindent [37] Jackson, J. D., ``Surface charges on circuit wires
and resistors play three roles," {\it Am. J. Phys.}, Vol. 64,
855-870, 1996.

\noindent [38] Shishkin, G. G., A. G. Shishkin, A. G. Smirnov, A. V.
Dudarev, A. V. Barkov, P. P. Zagnetov and Yu. M. Rybin,
``Investigation of possible electric potential arising from a
constant current through a superconductor coil ," {\it J. Phys. D}
Vol. 35, 497-502, 2002.

\noindent [39] Red\v zi\' c, D. V.,  ``Conductors moving in magnetic
fields: approach to equilibrium," {\it Eur. J. Phys.}, Vol. 25,
623-632, 2004.

\noindent [40] Red\v zi\' c, D. V., {\it Recurrent Topics in Special
Relativity: Seven Essays on the Electrodynamics of Moving Bodies},
authorial edition, Belgrade, 2006.

\noindent [41]Red\v zi\' c, D. V., ``Electromagnetostatic charges
and fields in a rotating conducting sphere," {\it Progress In
Electromagnetics Research}, Vol. 110, 383-401, 2010.

\noindent [42] O'Rahilly, A., {\it Electromagnetic Theory}, vol. 2,
589, Dover, New York, 1981.

\noindent [43] Miller, A. I., {\it Albert Einstein's Special Theory
of Relativity: Emergence (1905) and Early Interpretation
(1905-1911)}, 165-166, Springer, New York, NY, 1998.

\noindent [44] Lorrain, P., D. R. Corson and F. Lorrain, {\it
Electromagnetic Fields and Waves}, 3rd ed., 287-288, Freeman, New
York, NY, 1988.

\noindent [45] Ohanian, H. C., ``On the approach to electro- and
magneto-static equilibrium," {\it Am. J. Phys.}, Vol. 51,1020-1022,
1983.

\noindent [46] Gabuzda, D. C., ``Magnetic force due to a
current-carrying wire: a paradox and its resolution," {\it Am. J.
Phys.}, Vol. 55, 420-422, 1987.

\noindent [47] Brill, M. H., D. Dameron and T. E. Phipps, Jr., ``A
misapprehension concerning electric current neutrality ," {\it Phys.
Essays}, Vol. 24, 325-326, 2011.

\noindent [48] Folman, R, ``On the possibility of a relativistic
correction to the E and B fields around a current-carrying wire"
arXiv: 1109.2586v2, 2011.

\end {document}